\documentclass[10pt,twocolumn,letterpaper]{article}

\usepackage{cvpr}              

\usepackage[dvipsnames]{xcolor}

\definecolor{cvprblue}{rgb}{0.21,0.49,0.74}
\usepackage[pagebackref,breaklinks,colorlinks,citecolor=cvprblue]{hyperref}
\usepackage{newfloat}
\usepackage{listings}

\usepackage{booktabs}
\usepackage{threeparttable}
\usepackage{xcolor,colortbl}
\usepackage{amssymb}
\usepackage{amsmath}
\usepackage{multicol}
\usepackage{multirow}
\usepackage{tabularx}
\usepackage{pifont}
\usepackage{algorithm}
\usepackage{algorithmic}
\usepackage{makecell}
\usepackage{graphicx}

\newcommand{\method}{HEQuant}
\newcommand{\bm}{\boldsymbol}
\newcommand{\Dec}{\mathrm{Dec}}
\newcommand{\Enc}{\mathrm{Enc}}
\def\paperID{4034} 
\def\confName{CVPR}
\def\confYear{2024}
\definecolor{Gray}{gray}{0.85}
\title{\method: Marrying Homomorphic Encryption and Quantization for Communication-Efficient Private Inference}

\author{Tianshi Xu$^{1}$, Meng Li$^{1,2}$,Runsheng Wang$^{1}$\\
$^1$ School of Integrated Circuits, Peking University\\
$^2$ Institute for Artificial Intelligence, Peking University\\
{\tt\small meng.li@pku.edu.cn}
}
\begin{document}
\maketitle
\begin{abstract}


    Secure two-party computation with homomorphic encryption (HE) protects data
    privacy with a formal security guarantee but suffers from high
    communication overhead. While previous works, e.g., Cheetah, Iron, etc,
    have proposed efficient HE-based protocols for different neural network
    (NN) operations, they still assume high precision, e.g., fixed point 37 bit,
    for the NN operations and ignore NNs' native robustness against
    quantization error. In this paper, we propose \method, which features
    low-precision-quantization-aware optimization for the HE-based protocols.
    We observe the benefit of a naive combination of
    quantization and HE quickly saturates as bit precision goes down. Hence, to
    further improve communication efficiency, we propose a series of
    optimizations, including an intra-coefficient packing algorithm and
    a quantization-aware tiling algorithm, to simultaneously reduce the number
    and precision of the transferred data. Compared with prior-art HE-based
    protocols, e.g., CrypTFlow2, Cheetah, Iron, etc, \method~achieves
    $3.5\sim 23.4\times$ communication reduction and $3.0\sim 9.3\times$ 
    latency reduction. Meanwhile, when compared with prior-art
    network optimization frameworks, e.g., SENet, SNL, etc, \method~also achieves
    $3.1\sim 3.6\times$ communication reduction.
    
\end{abstract}
    
\section{Introduction}
\label{intro}


\begin{figure}[!tb]
    \centering
    \includegraphics[width=1.0\linewidth]{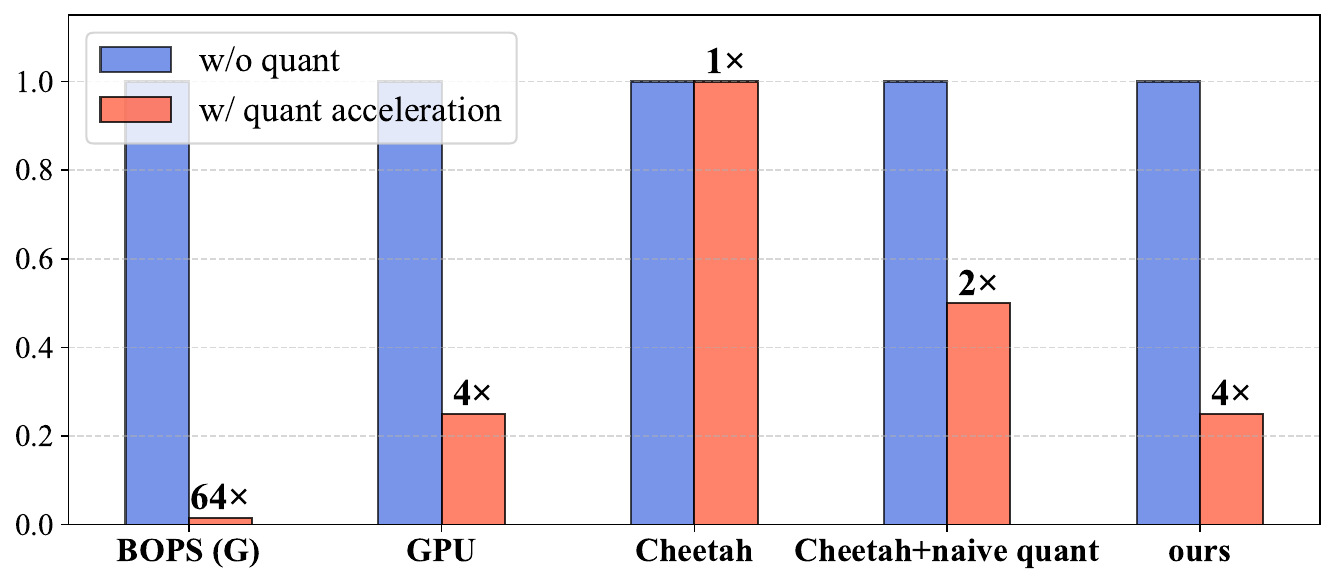}
    \caption{The acceleration achieved through network quantization (from 32-bit to 4-bit) on Bit Operations (BOPS), GPU, prior-art 2PC frameworks Cheetah and Cheetah with naive quant as well as our proposed framework.}
    \label{fig:intro_1}
\end{figure}

The last decade has witnessed the rapid evolution of deep learning (DL) as well as its increasing adoption in sensitive and private applications,
including face authentication \cite{azouji2022efficientmask_face}, medical diagnosis \cite{kaissis2021end_medical}, personal assistant \cite{brill2022siri}, etc.
Privacy emerges as a major concern and leads to a growing demand for privacy-preserving DL 
\cite{Choi_Reagen_Wei_Brooks_Impala_2022,Demmler_Schneider_ABY_2015,Gupta_Kumaraswamy_Chandran_Gupta_LLAMA_2022,kumar2020cryptflow}.

Secure two-party computation (2PC) based on homomorphic encryption (HE) has recently been proposed and attracted a lot of attention 
\cite{huang2022cheetah,Dathathri_Saarikivi_Chen_Laine_Lauter_Maleki_Musuvathi_CHET_2019,Kim_Park_Kim_Kim_Ahn_HyPHEN_2023}. 
It helps to solve the dilemma between the DL model owner and the data owner:
while the two parties want to jointly apply the model to the data, they do not want to give out the model or data directly. 
2PC enables the computation and permits cryptographically-strong privacy protection \cite{hao2022iron,Juvekar_Vaikuntanathan_gazelle_2018,gilad2016cryptonets,kumar2020cryptflow}.

However, the privacy protection of the 2PC-based inference is achieved at the cost of high communication complexity since numerous high-precision ciphertexts (e.g., 109-bit ciphertext for Cheetah \cite{huang2022cheetah}) are transferred between the two parties.
Previous works, e.g., Cheetah \cite{huang2022cheetah}, Iron \cite{hao2022iron}, Falcon \cite{xu2023falcon}, etc, have proposed efficient HE-based
protocols for representative deep neural network (DNN) operations like convolutions, matrix multiplications, etc.
However, they all assume high-precision plaintext operations (e.g., 37-bit plaintext for Cheetah \cite{huang2022cheetah}) are required
and ignore the native robustness of deep neural networks (DNNs) towards quantization error \cite{gholami2022survey_quant}. 
As shown in Figure~\ref{fig:intro_1}, prior works have shown that 4-bit quantization can achieve plausible accuracy with
significant compute reduction \cite{liu2022nonuniform,yamamoto2021LCQ,bhalgat2020lsq+,yao2021hawq,dong2020hawq},
existing protocols like Cheetah cannot leverage such property for communication reduction.

While a naive combination of quantization and HE can help reduce the communication as shown in Figure~\ref{fig:intro_1},
we find further efficiency improvement is challenging: on one hand, to ensure correct decryption, the plaintext modulus cannot
be reduced arbitrarily, leading to saturated efficiency gain for low-precision quantized networks, e.g., 4-bit or below;
on the other hand, the input and output communication for a convolution layer is highly imbalanced while how to further reduce
the output communication remains unclear.

In this paper, we propose a HE-based 2PC framework that is optimized for low-precision quantized DNN inference, named \method.
\method~features two main techniques to fully leverage the DNN quantization. First, \method~proposes an intra-coefficient
packing algorithm to pack multiple input elements into each coefficient and allows further communication reduction given
the saturated plaintext modulus. Moreover, \method~further proposes a novel quantization-aware tiling strategy for linear 
operations that significantly reduces output communication. Our main contributions can be summarized as follows:
\begin{itemize}
    \item We observe existing HE-based 2PC frameworks are not optimized for quantization and propose \method, the first 2PC 
        framework that marries quantization and HE for efficient private inference.
    \item We propose an intra-coefficient packing algorithm and a quantization-aware tiling algorithm to further improve the
        communication efficiency for low-precision DNNs.
    \item \method~outperforms state-of-the-art (SOTA) HE-based 2PC frameworks, including CrypTFlow2, Cheetah and Falcon
        with $3.5\sim 23.4\times$ communication reduction and $3.0\sim 9.3\times$ latency reduction on CIFAR-100, Tinyimagenet 
        and ImageNet datasets. Compared with prior-art network optimization algorithms, including  
        ReLU-optimized networks and network pruning methods, \method~achieves $3.5\times$ communication reduction
        and $4.3\times$ latency reduction on average with higher accuracy.
\end{itemize}

\section{Preliminaries}


\begin{table}[!tb]
    \centering
    \caption{Notations used in the paper.}\label{tab:notation}
    \resizebox{1.0\linewidth}{!}{
        \begin{tabular}{c|c}
        \toprule
        Notations & Meanings \\
        \midrule
        $\lceil \cdot \rceil$,$\lfloor\cdot \rfloor$,$\lfloor\cdot \rceil$ & Ceiling, flooring, and rounding operations \\
        \hline
        $\Enc(\cdot), \Dec(\cdot)$ & Homomorphic encryption and decryption\\
        \hline
        $\boxplus,\boxminus,\boxtimes $  & Homomorphic addition, subtraction and multiplication  \\
        \hline
        $P, Q$ & Plaintext modulus, and ciphertext modulus \\
        \hline
        \multirow{2}{*}{$N, p,q$} & Polynomial degree, bit width of plaintext \\
                               & ($p = \lceil \log_2 P \rceil$) and ciphertext ($q = \lceil \log_2 Q \rceil$) \\
        \hline
        $b_x,b_w,b_{acc}$ & Bit width of activation, weight, and accumulation\\
        \hline
        $H, W, C$ & Height, width, and channels of Conv2D input tensor \\
        \hline
        $R, K$ & Kernel size and the number of kernels of Conv2D filters \\
        \bottomrule
        \end{tabular}
    }
\end{table}


\subsection{Threat Model}
\label{subsec:threat_model}


We focus on private DNN inference involving two parties,
i.e., a server and a client. The server holds the model with private weights, 
while the client holds private inputs~\cite{huang2022cheetah,hao2022iron,xu2023falcon}. The model architecture, including
the number of layers as well as the types, dimensions, and bit widths for each
layer, is publicly known to both parties~\cite{Mishra_Delphi_2020,rathee2020cryptflow2}.
At the end of the protocol execution, the client acquires the inference results 
without leaking any information to the server. Consistent with previous works~\cite{gilad2016cryptonets,hussain2021coinn,rathee2021sirnn,mohassel2017secureml}, we adopt an
\textit{honest-but-curious} security model where both parties adhere to the protocol's specifications but also attempt to learn more than permitted.
Hence, malicious attacks like adversarial attacks \cite{lehmkuhl2021muse} are not the focus of the paper.

\subsection{HE-based 2PC Inference}
\begin{table}[!tb]
    \centering
    \caption{Comparison with previous works. The Comm. is measured by Conv. with dimension (H,W,C,K,R)=(14,14,256,256,3) and ReLU with input dimension (14,14,256).}
    \label{tab:relat_comp}
    \resizebox{1.0\linewidth}{!}{
    \begin{tabular}{c|cccc}
    \toprule 
    \multirow{1}{*}{Framework}  &  \multicolumn{1}{c}{Protocol Opt.}&Network Opt. & Conv. Comm. & ReLU Comm.  \\
    \midrule
    \cite{cho2022SNL,jha2021deepreduce,kundu2023SENet,cho2022sphynx}  & - & ReLU Reduction &-&$0.6\sim 1$ MB \\
    \hline
    \cite{liu2019metapruning}  & - & Network Pruning &9 MB& 2 MB \\
    \hline
    SiRNN\cite{rathee2021sirnn} & \makecell{Quantized MatMul \\ Bit Extension/Trunc} & \makecell{16 bit Weight/Act.\\40+ bit Accum.} &9079 MB&15 MB \\
    \hline
    Cheetah\cite{huang2022cheetah}  & \makecell{HE Coeff.Pack for Conv/MatMul\\Output Coeff. Optim.\\ReLU Opt.}    & 37 bit Quant. &12 MB &3 MB  \\
    \hline
    Iron\cite{hao2022iron} & \makecell{Tile for MatMul\\Softmax/GeLU Opt.}& 37 bit Quant.  &14 MB &3 MB\\
    \hline
    \makecell{\method \\ (ours)}       &  \makecell{Intra-Coeff.Pack for Conv/MatMul \\
    Tile for Conv \\Output Coeff. Optim. }& \makecell{Low-Precision Quant.\\e.g. 4/3 bit}&4 MB &0.2 MB \\
    \bottomrule
    \end{tabular}
    }
\end{table}

\textbf{Related works} There are often two categories of HE-based 2PC frameworks, including the end-to-end HE-based frameworks
\cite{gilad2016cryptonets,lee2022privacy,lee2022low,park2023toward_fhe,kim2023optimized_fhe,fan2023tensorfhe,kim2022ark,onoufriou2021fully}
and the hybrid frameworks that combine both arithmetic secret sharing (ASS) and HE
\cite{Liu_Juuti_MiniONN_2017,Mishra_Delphi_2020,Garimella_Ghodsi_Jha_Garg_Reagen_2022,lu2023bumblebee,kim2023hyphen}. 
We focus on the second category as it supports different non-linear activations more accurately 
\cite{huang2022cheetah,Juvekar_Vaikuntanathan_gazelle_2018,hao2022iron,xu2023falcon,rathee2020cryptflow2}. We compare \method~with existing works including 2PC frameworks\cite{rathee2021sirnn,huang2022cheetah,hao2022iron} and network optimization works\cite{cho2022SNL,jha2021deepreduce,kundu2023SENet,cho2022sphynx,liu2019metapruning}, as shown in Table~\ref{tab:relat_comp}.
\method~leverages an intra-coefficient packing for convolution with tiling and output coefficient optimization. Furthermore, \method~utilizes low-precision network quantization and significantly reduces both linear and non-linear layers' communication. In addition, we summarize the notations used in the paper in Table~\ref{tab:notation}.

\begin{figure}[!tb]
    \centering
    \includegraphics[width=1.0\linewidth]{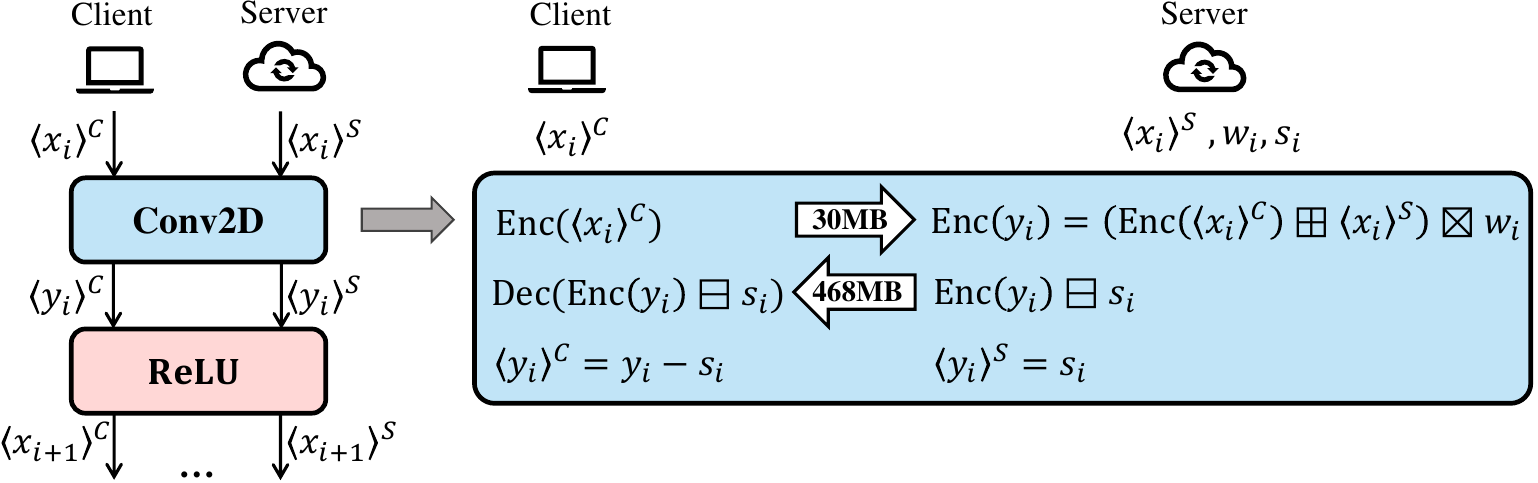}
    \caption{DNN private inference based on HE (linear layer). 
    }
    \label{fig:pre_HE}
\end{figure}

\noindent \textbf{The flow of 2PC-based inference} Figure~\ref{fig:pre_HE} shows the flow of 2PC-based inference. With ASS, each intermediate activation
tensor, e.g., $x_i$, is additively shared where the server holds $\langle x_i\rangle^S$ and the client holds $\langle x_i\rangle^C$ such that $x_i = \langle x_i\rangle^S+\langle x_i\rangle^C \mod P$ \cite{mohassel2017secureml}.
For a linear layer, e.g. Conv2D, to compute the result of $y_i$ without revealing both parties' information, the client first encrypts its share as $\Enc(\langle x_i \rangle^C)$ and sends it to the server.
Since HE supports homomorphic addition and multiplication without the need for decryption,
the server can compute $\Enc(y_i)$ and blind $y_i$ by subtracting a randomly sampled $s_i$.
Finally, the server sends the result to the client for decryption.
The server and client now obtain the secret shares of the Conv2D output
$y_i$\footnote{Note it is possible to move the HE computation to the pre-processing stage~\cite{Mishra_Delphi_2020}, the required total communication remains almost the same.}.
For a non-linear layer, e.g. ReLU, we omit the detailed protocols and refer interested readers to~\cite{rathee2020cryptflow2,rathee2021sirnn,huang2022cheetah}. 

\begin{figure}[!tb]
    \centering
    \includegraphics[width=0.87\linewidth]{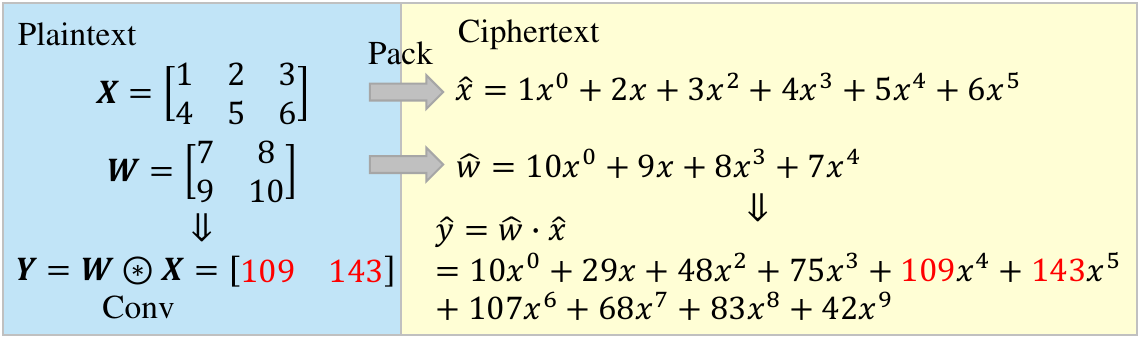}
    \caption{A convolution example of coefficient packing.
    }
    \label{fig:coe_pack}
\end{figure}

At a lower level, HE encrypts the plaintexts into ciphertext polynomials and computes over the polynomials, which can be regarded as 1-dimensional vectors.
In contrast, DNN processes high-dimensional tensors.
Therefore, packing becomes necessary to encode tensors into polynomials. Cheetah~\cite{huang2022cheetah} identifies that polynomial multiplications natively perform convolution operations and introduces a polynomial coefficient packing method. In this approach, elements in tensors are encoded into coefficients, and the correct result can be extracted from the outcome of polynomial multiplication. A toy example is illustrated in Figure~\ref{fig:coe_pack}.


\subsection{Communication Complexity of HE-based 2PC}
\label{subsec:comm_complexity}

\renewcommand\arraystretch{1.5}
\begin{table}[!tb]
    \centering
    \caption{Comm. complexity of HE-based 2PC frameworks.}\label{tab:comm_complexity}
    \begin{threeparttable}
    \resizebox{\linewidth}{!}{
        \begin{tabular}{l|cc|c}
        \toprule
        \multirow{2}{*}{\large{Layer Type}}&\multicolumn{2}{c|}{\large{Linear Layer (Conv.)}}  &\multirow{2}{*}{\large{Non-linear Layer}} \\
        \cline{2-3}
        &\large{Input Ciphertext} & \large{Output Ciphertext}&  \\
        \midrule 
        \large{Comm. Complexity}&\large{$\lceil \frac{HWC}{N}\rceil \times Nq$} & \large{$( \lceil \frac{HWCK}{N}\rceil  \times N +HWK)q$}&{\large{$\mathrm{O}(b_x)$}} \\
        \bottomrule 
        \end{tabular}
    }
    \end{threeparttable}
\end{table}
\renewcommand\arraystretch{1}

The communication of a linear layer is mainly consumed by the input and output ciphertext transfer.
The communication can be accurately estimated by the product of the number of ciphertext polynomials and the size of each polynomial.
For each polynomial, the communication is the product of the polynomial degree and the bit width of each coefficient.
Hence, the communication of the input ciphertexts can be estimated as $\#Poly \times Nq$.
For the output ciphertext, because each activation is encrypted into a scalar $b$ and a polynomial $\bm{a}$,
extra scalars need to be transferred. Hence, following Cheetah \cite{huang2022cheetah}, the output communication can be estimated
with $\#Poly \times Nq + nq$, where $n$ is the number of output activations, $\#Poly \times Nq$ is the communication of $\bm{a}$,
and $nq$ is the communication of $b$.
For a nonlinear layer like ReLU, the communication scales with the bit width of activations $b_a$~\cite{rathee2021sirnn}.
We summarize the communication complexity in Table~\ref{tab:comm_complexity}.

\noindent \textbf{Operator tiling for linear layer} 
When the input tensor size gets larger than the polynomial degree, it needs to be tiled into a group of small tensors before encoding into polynomials.
Operator tiling determines the strategy of splitting the input tensors and directly impacts the communication cost \cite{hao2022iron}.
For a general convolution, previous works~\cite{hao2022iron,xu2023falcon} focus on minimizing the
communication of input polynomials, i.e., they try to pack as many input activations as possible.
Hence, following Cheetah, given $N$, the numbers of input polynomials and output polynomials are $\lceil \frac{HWC}{N}\rceil$ and $\lceil \frac{HWCK}{N}\rceil$, respectively \cite{huang2022cheetah}.

\noindent \textbf{Selection of HE parameters $p$ and $q$} In HE, $p$ represents the bit width of the plaintext field. To avoid overflow,
the lower bound of $p$ is the required accumulation bit width for a linear layer, i.e., $b_{acc}$. Given $p$,
$q-p$ determines the noise budget of the HE-based computation, which is determined by the depth of multiplications.
In addition, we introduce the background of network quantization in Appendix~\ref{app:quantization}.

\section{Motivation}
\label{sec:Motivation}

\begin{figure}[!tb]
    \centering
    \includegraphics[width=0.86\linewidth]{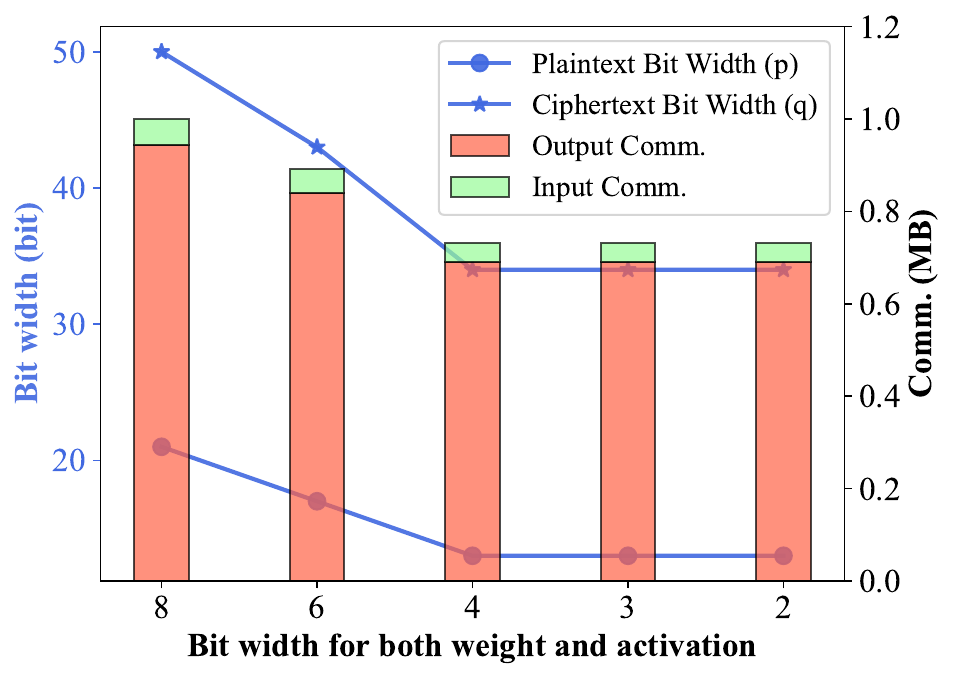}
    \caption{The input and output communication as well as the bit width of plaintext and ciphertext for different bit precision quantization.}
    \label{fig:moti_pq}
\end{figure}

\begin{figure*}[!tb]
    \centering
    \includegraphics[width=0.84\linewidth]{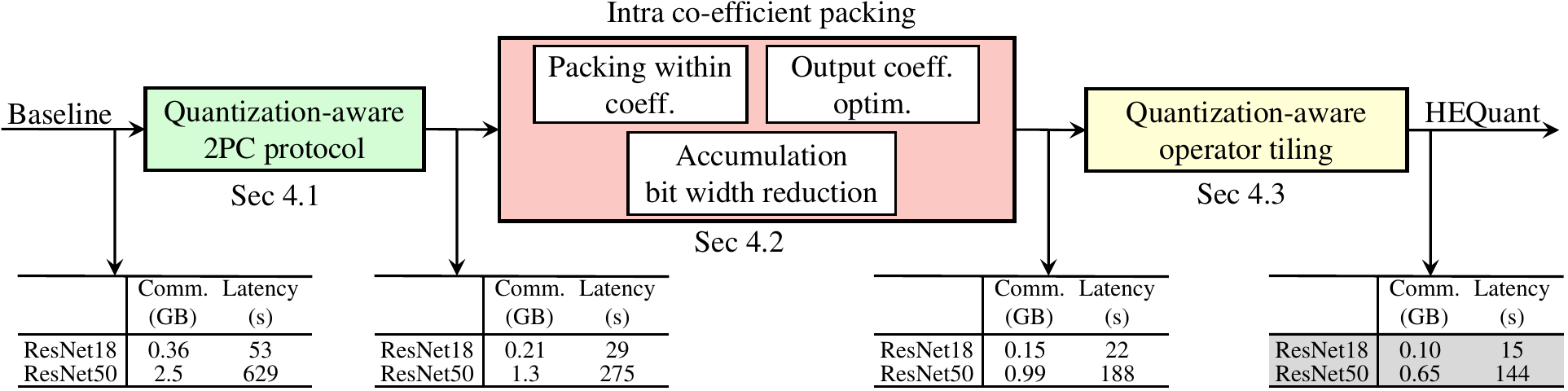}
    \caption{Overview of \method~and the communication/latency cost after each optimization step. The examples are ResNet18 on CIFAR-100 and ResNet50 on ImageNet. Coeff. represents coefficient. 
    }
    \label{fig:overview}
\end{figure*}

In this section, we discuss the key observations that motivate our quantized HE-based 2PC framework \method.
We will focus on the linear layers as they contribute to the majority of the communication\cite{huang2022cheetah,hao2022iron,lu2023bumblebee}.

\textbf{Observation 1: both HE parameters $p$ and $q$ can be reduced for low-precision quantized DNN linear operators,
leading to better communication efficiency.} 
As introduced in Section~\ref{subsec:comm_complexity}, for a linear operator, $p$ is determined by
the accumulation bit precision while $q - p$ is determined by the required noise budget and is roughly
fixed for a given operator.
Low-precision quantization helps to directly reduce the required accumulation bit precision,
and thus, improves the communication efficiency. The observation is also verified in Figure~\ref{fig:moti_pq}.
Considering the robustness of DNNs to quantization error \cite{gholami2022survey_quant,nagel2021white_quant,krishnamoorthi2018quantizing_white}, low-precision quantization becomes a promising
technique to improve the efficiency of the HE-based 2PC framework.

\textbf{Observation 2: while networks continue to be quantized for sub-4-bit precision, $p$ and $q$ cannot be further 
reduced, leading to saturated communication efficiency improvement.} Previous research has demonstrated 
the feasibility of quantizing networks to sub-4 bits, such as 2-bit or even 1-bit, 
while maintaining comparable accuracy \cite{qin2020forward,qin2020binary,liu2022nonuniform,yuan2023comprehensive}. 
However, as shown in Figure~\ref{fig:moti_pq}, we empirically find $p$ cannot be reduced below $13$ bits
to ensure correct computation, leading to saturated communication reduction, especially for sub-4-bit
quantization. Hence, how to further improve the communication efficiency for ultra-low precision quantized DNNs
is also an important question to answer.


\textbf{Observation 3: the input and output communication are not balanced.} As shown in Figure~\ref{fig:moti_pq},
regardless of the bit precision, the communication of output ciphertexts is much larger compared to the input
ciphertexts as more output ciphertexts need to be transferred. Table~\ref{tab:comm_complexity} also theoretically supports it.
While past research has proposed to tile the input for matrix multiplications and depthwise convolutions
to reduce the number of output polynomials \cite{hao2022iron,xu2023falcon},
a general tiling algorithm has not yet been proposed for convolutions especially when considering the impact of quantized DNNs.
Moreover, as introduced in Section~\ref{subsec:comm_complexity}, another way to reduce output communication is to lower the 
coefficient bit widths. Hence, how to reduce the coefficient bit precision beyond $q$ without impacting 
the output correctness is also an important direction to study.

\section{\method: Communication-Efficient 2PC Framework}

In this section, we present \method, an efficient 2PC framework based on quantization-aware HE protocols.
The overview of \method~is shown in Figure~\ref{fig:overview}. Based on Observation 1, given a quantized
network, we first propose a quantization-aware 2PC protocol and optimize the communication by reducing 
the HE parameters $p$ and $q$, achieving more than $2\times$ latency reduction (Section~\ref{subsec:quant}).
Then, to further improve the efficiency, we propose intra-coefficient packing to allow packing multiple
low-precision activations into a single coefficient to reduce the number of input and output ciphertexts.
A series of bit width optimizations for both the plaintext and ciphertext, i.e., $p$ and $q$, are also 
developed to enable dense packing. The proposed optimization further
enables $1.4\times$ latency reduction (Section~\ref{subsec:intra}). In the last step, a quantization-aware
tiling algorithm is proposed for general linear operations (Section~\ref{subsec:tiling}). With 3-bit weight and 3-bit
activation, \method~achieves almost $4\times$ latency reduction over prior-art Cheetah framework \cite{huang2022cheetah}.



\subsection{HE-based quantized private inference}
\label{subsec:quant}

\begin{figure}[!tb]
    \centering
    \includegraphics[width=0.75\linewidth]{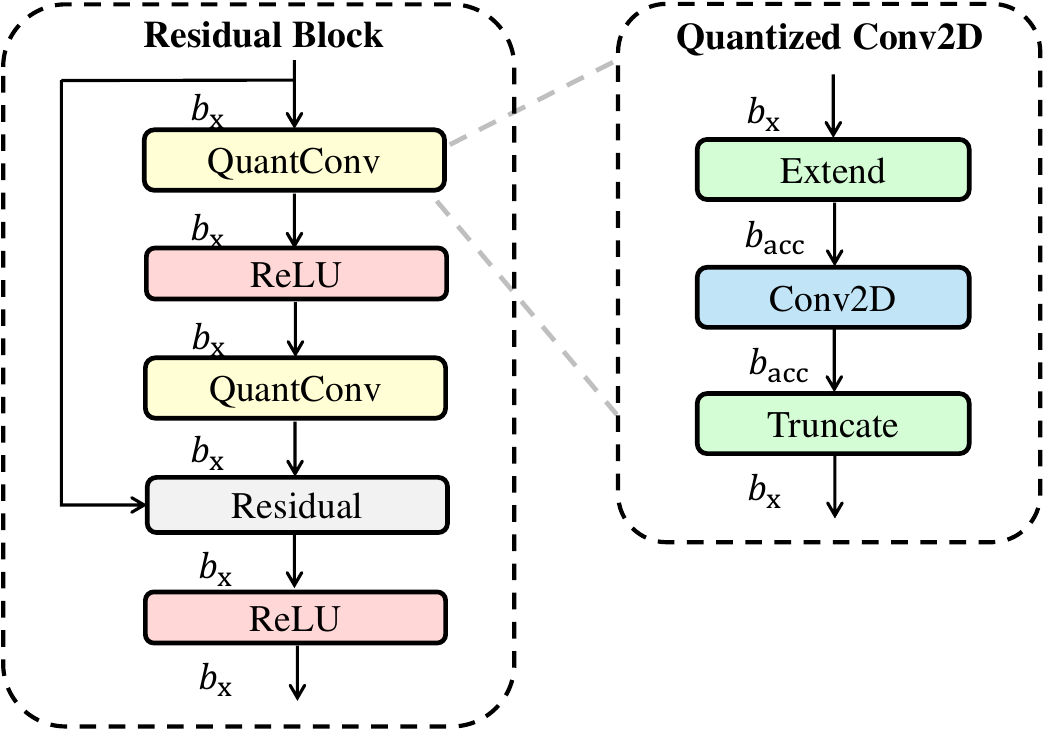}
    \caption{Quantized HE-based 2PC frameworks on a residual block. b\_ means the bit width for activation.
    }
    \label{fig:pipeline}
\end{figure}

Based on the observations discussed in Section~\ref{sec:Motivation}, We propose the quantized HE-based 2PC frameworks in Figure~\ref{fig:pipeline}. 
We take a residual block as an example. To avoid overflow, the HE-based convolution protocol requires the plaintext bit width $p \geq b_{acc}$.
However, the convolution input is $b_a$-bit.
Hence, bit extension protocols are needed to increase the bit width of activations from $b_a$-bit to $b_{acc}$-bit.
After the convolution, we truncate output activations from $b_{acc}$-bit to $b_a$-bit to conduct low-precision ReLU to reduce communication since the communication
of ReLU scales linearly with the activation bit width as shown in Table~\ref{tab:comm_complexity}.
The bit width extension, truncation, and residual protocols are all implemented following previous work SiRNN~\cite{rathee2021sirnn}.
We refer interested readers to \cite{rathee2021sirnn} and focus on describing the quantization-aware HE-based convolution protocol in the following sections.

\subsection{Intra Co-efficient Packing}
\label{subsec:intra}
\begin{figure}[!tb]
    \centering
    \includegraphics[width=0.91\linewidth]{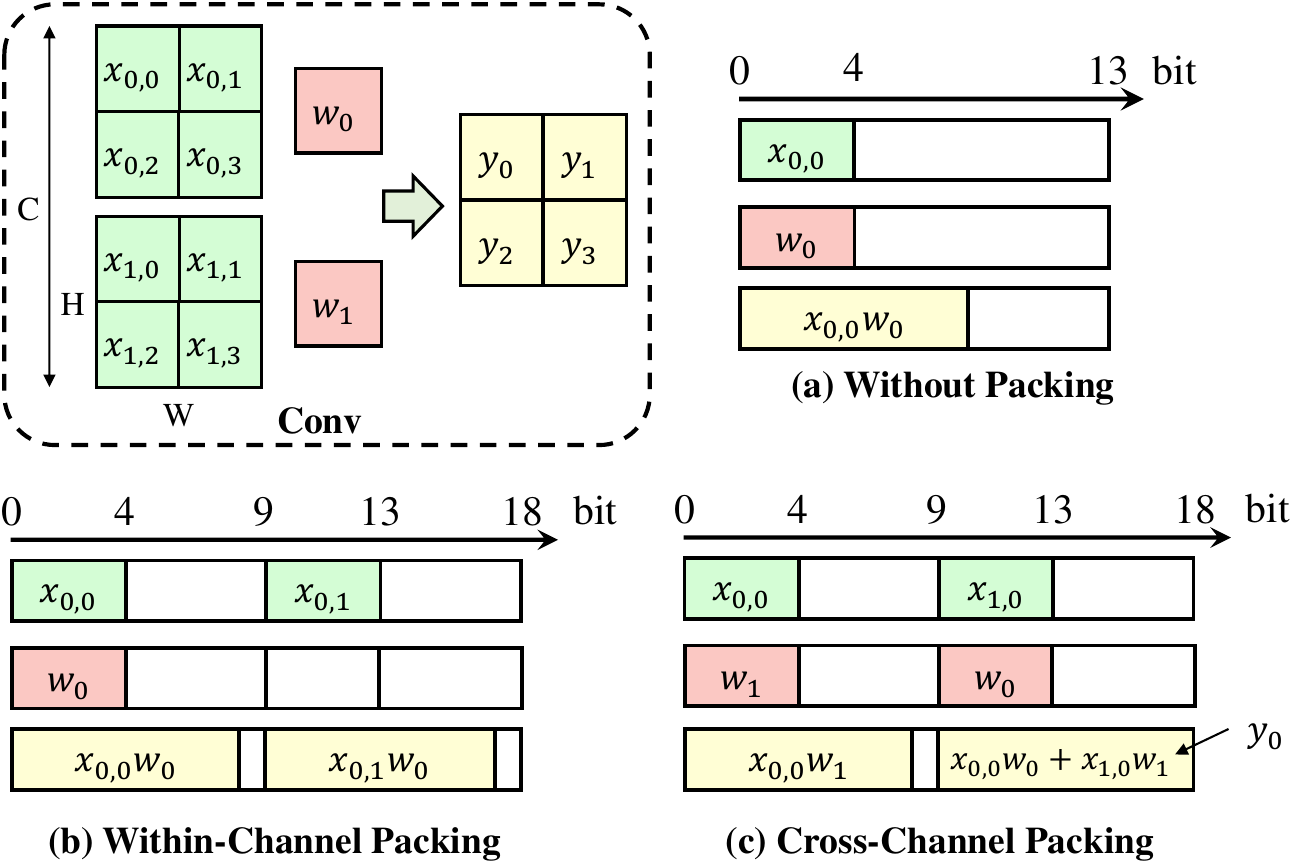}
    \caption{Examples of the proposed intra coefficient packing.
    }
    \label{fig:pack1}
\end{figure}


\noindent \textbf{Packing within coefficient}
Following Observation 2 in Section~\ref{sec:Motivation}, 
the communication reduction saturates for sub-4-bit quantized linear layers due to the lower bound of $p$ and $q$.
We observe although $p$ and $q$ cannot be further reduced, it is possible to pack multiple low-precision activations into a single coefficient.
Consider the example in Figure~\ref{fig:pack1}. Two 4-bit values are packed into one 18-bit coefficient
with the two values occupying $0-4$ and $9-13$ bits, respectively. Such packing strategy enables us to
correctly compute one multiplication with a 4-bit weight and one accumulation (which at most requires $9$-bit accumulation)
while reducing the number of coefficients by $2\times$.
Based on the observation, for a convolution, we propose two strategies for intra-coefficient packing, namely Within-Channel Packing and Cross-Channel Packing.

 \textbf{Within-Channel Packing}
As shown in Figure~\ref{fig:pack1} (b), Within-Channel Packing combines two activations
within the same channel into one coefficient. Since these two activations need to multiply the same weight, the
weight packing stays the same. After the homomorphic multiplication, each coefficient will also have two output
activations, resulting in a 2$\times$ reduction in the input and output polynomials as well as the number
of output coefficients.


\textbf{Cross-Channel Packing}
Figure~\ref{fig:pack1} (c) shows the Cross-Channel Packing strategy. Two activations
along the same channel are now packed into the same coefficient. Since the two activations need to multiply
different weights and the products need to be added together, the weights packing needs to be modified accordingly. After the homomorphic multiplication, each coefficient only contains one output
activation in higher bit locations. Hence, Cross-Channel Packing reduces the input and output polynomials by $2\times$
while the number of output coefficients remains unchanged.

\renewcommand\arraystretch{1.5}
\begin{table}[!tb]
    \centering
    \caption{Comparsion of packing methods on communication and limitation as well as the measured communication for a Conv. with dimension (H,W,C,K,R)=(14,14,32,32,1), $b_{acc}=8$. ``+'' represents output coefficient optimization. The actual value for p and q can be different for Cheetah and our method.
    }\label{tab:cmp_pack}
    \resizebox{1.0\linewidth}{!}{
        \begin{tabular}{c|cccc}
        \toprule 
        \large{Packing Method}  &  \large{Input Comm.} & \large{Output Comm. } & \large{Conv. Comm. } & \large{Limitation}  \\
        \midrule
        \large{Cheetah}      & \large{$\lceil \frac{HWC}{N}\rceil  \times Nq$}  &\large{$(\lceil \frac{HWCK}{N}\rceil\times N+HWK)q$} &\large{0.90 MB} & - \\
        \midrule
        \large{Within-Channel}      & \large{$\lceil \frac{HWC}{2N}\rceil  \times Nq$}  &\large{$(\lceil \frac{HWCK}{2N}\rceil\times N+\frac{HWK}{2})q$} &\large{0.51 MB}& \large{kernel size=1}  \\
        \large{Cross-Channel}      &\large{$\lceil \frac{HWC}{2N}\rceil  \times Nq$}   &\large{$(\lceil \frac{HWCK}{2N}\rceil\times N+HWK)q$}&\large{0.52 MB}& \large{/}\\
        \hline
        \large{Within-Channel+}      & \large{$\lceil \frac{HWC}{2N}\rceil  \times Nq$}  &\large{$\lceil \frac{HWCK}{2N}\rceil\times N(p+\sigma)+\frac{HWKp}{2}$} &\large{0.33 MB}& \large{kernel size=1}  \\
        \large{Cross-Channel+}      &\large{$\lceil \frac{HWC}{2N}\rceil  \times Nq$}   &\large{$\lceil \frac{HWCK}{2N}\rceil\times N(\frac{p}{2}+\sigma)+HWKp$}&\large{0.23 MB}& \large{1-bit error}\\
        \bottomrule
        \end{tabular}
    }
\end{table}
\renewcommand\arraystretch{1}


\textbf{Comparison}
These two packing schemes have different advantages and disadvantages and are suitable for different situations. 
For one thing, Cross-Channel Packing can be applied to convolution with any kernel size while Within-Channel Packing can only be used for convolution with kernel size 1. 
For another, they also have different communication impacts as summarized in Table~\ref{tab:cmp_pack}. 
As can be observed, Within-Channel Packing achieves smaller output communication due to fewer output activations.
Additionally, we observe that in Cross-Channel Packing, only the higher bit locations of each output activation are useful.
This indicates opportunities to further optimize the communication as will be discussed later.

\begin{figure}[!tb]
    \centering
    \includegraphics[width=0.9\linewidth]{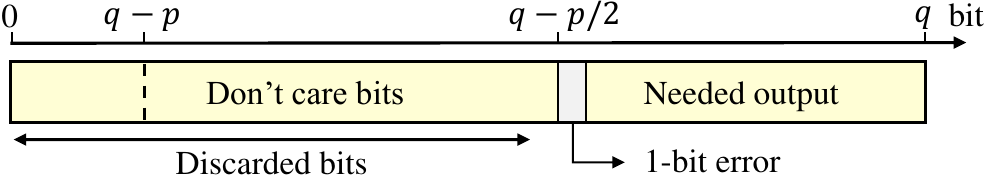}
    \caption{Output coefficient optim. by discarding low-end bits.
    }
    \label{fig:hack_optim}
\end{figure}

\noindent \textbf{Output coefficient optimization}
To further reduce the communication of output ciphertext, we further reduce the coefficient bit widths of output ciphertexts.
Our bit width optimization is based on the following two observations.
Firstly, recall from Section~\ref{sec:Motivation}, a $p$-bit output $m$ is encrypted into a scalar $b$ and a polynomial $\bm{a}$.
During decryption, we compute $m=\lfloor (b+\bm{a}^\text{T}\bm{s})P/Q\rfloor$ to reconstruct the output $m$,
where $Q$ and $P$ are the ciphertext modulus and plaintext modulus, respectively. 
Hence, the low $q - p$ bits of each coefficient may not be fully needed.
Secondly, as mentioned above, in our Cross-Channel Packing method, only the higher bit locations of each coefficient are useful
while the other bits may not be fully required for decryption.
We illustrate the observations in Figure~\ref{fig:hack_optim}.

Based on the observations, we propose to discard the lower bits before ciphertext transmission. We observe discarding these lower bits
is equivalent to truncating coefficients and may introduce a single-bit error after truncation. 
The error probability depends on the number of discarded bits. Moreover, since each output activation is encrypted into a scalar $b$
and a polynomial $\bm{a}$, how to discard $b$ or $\bm{a}$ will lead to different error rate and communication.
We empirically decide to skip the low $\lfloor \log_2(Q/P)\rceil-1$ of $b$ and the low $\lfloor \log_2(Q/P)\rceil +\frac{p}{2}-1-\sigma$ of $\bm{a}$
during output ciphertext transmission, where $\sigma$ is the standard deviation of $\bm{a}^\text{T}\bm{s}$. 
Additional details are available in Appendix~\ref{app:hack} and in the experimental results, a comprehensive ablation study justifies this choice.

With the output coefficient optimization, the communication complexity of the two packing methods proposed above is shown in Table~\ref{tab:cmp_pack},
denoted as ``Within-Channel+'' and ``Cross-Channel+''.

\noindent \textbf{Accumulation bit width reduction}
As analyzed in Section~\ref{subsec:comm_complexity}, the plaintext bit width $p$ is selected according to the accumulation bit width $b_{acc}$.
Hence, how to reduce the accumulation bit width directly impacts the inference cost.
We adopt three strategies to estimate the required $b_{acc}$.

\textbf{Strategy 1} We directly set $b_{acc} = b_w+b_x+\log_2(e)$ to avoid any overflow, where $e$ is the number of accumulations. 
This is extensively used in previous works~\cite{rathee2021sirnn,bhalgat2020lsq+,yamamoto2021LCQ,liu2022nonuniform}.
However, we find that accumulated values rarely reach the theoretical upper bound as the weights can be either positive or negative
and most weights have small magnitude and do not reach $2^{b_w}$.
Hence, we propose Strategy 2 to reduce the accumulation bit width.



\textbf{Strategy 2} Considering the weights are frozen during inference, for one output channel,
we have $\sum_{i=1}^e |x_i w_i| \leq \sum_{i=1}^e |x_i| |w_i| \le 2^{b_x}\sum_{i=1}^e |w_i| = 2^{b_x}\left \| \bm{w} \right \|_1$.
So we just need $b_{acc} = b_x+\lceil \log_2(\left \| \bm{w} \right \|_1)\rceil $ to avoid overflow. 
A weakness is that it may leak $\lceil \log_2(\left \| \bm{w} \right \|_1) \rceil$. However, we find it's still hard for the client to obtain weights.

\textbf{Strategy 3} 
We further observe that most activations also cannot reach the upper bound of $2^{b_x}$.
Hence, we propose to leverage the training dataset to directly collect the accumulation bit width required for each layer. 
While this minimizes $b_{acc}$, it may introduce the overflow error if the testing dataset 
does not follow the same statistics as the training dataset.
We compare the three strategies in Section~\ref{subsec:ablation} and empirically find Strategy 3 does not impact the model testing accuracy. 
We also show the layer-wise accumulation bit width comparison of three strategies in Appendix~\ref{app:accum_reduce}.


\subsection{Quantization-aware operator tiling}
\label{subsec:tiling}

Based on Observation 3 in Section~\ref{sec:Motivation}, the input and output ciphertext communication are highly imbalanced,
which motivates previous works to tile the operator to minimize the communication. Specifically, Iron and Falcon design tiling
strategies for matrix multiplication and depthwise convolution, respectively.
However, we find there exist two main drawbacks in previous methods: \textbf{\underline{1)}} lack of tiling strategy for general convolution;
\textbf{\underline{2)}} do not consider the layerwise difference in $p, q$ induced by the network quantization and output coefficient optimization.

To address these issues, we propose a novel quantization-aware tiling algorithm for general convolution.
We first conduct a theoretical analysis of communication cost and formulate it as a nonlinear programming problem. Based on Section~\ref{subsec:comm_complexity}, when introducing tiling at the channel level, we use $C_x,C_w$ to denote the number of input and weight channels in each polynomial, respectively. Then the communication cost is given by the equation:
\begin{equation}\label{eq:comm_cost1}
    \lceil \frac{C}{C_x}\rceil\times Nq+(\lceil \frac{K}{C_w}\rceil\times N+HWK)q
\end{equation}
where $\lceil \frac{C}{C_x}\rceil$ and $\lceil \frac{K}{C_w}\rceil$ are the number of input and output polynomials, respectively. Furthermore, when output coefficient optimization is introduced, the communication cost of Equation~\ref{eq:comm_cost1} can be replaced by that in Table~\ref{tab:cmp_pack}. 

\begin{figure}[!tb]
    \centering
    \includegraphics[width=0.70\linewidth]{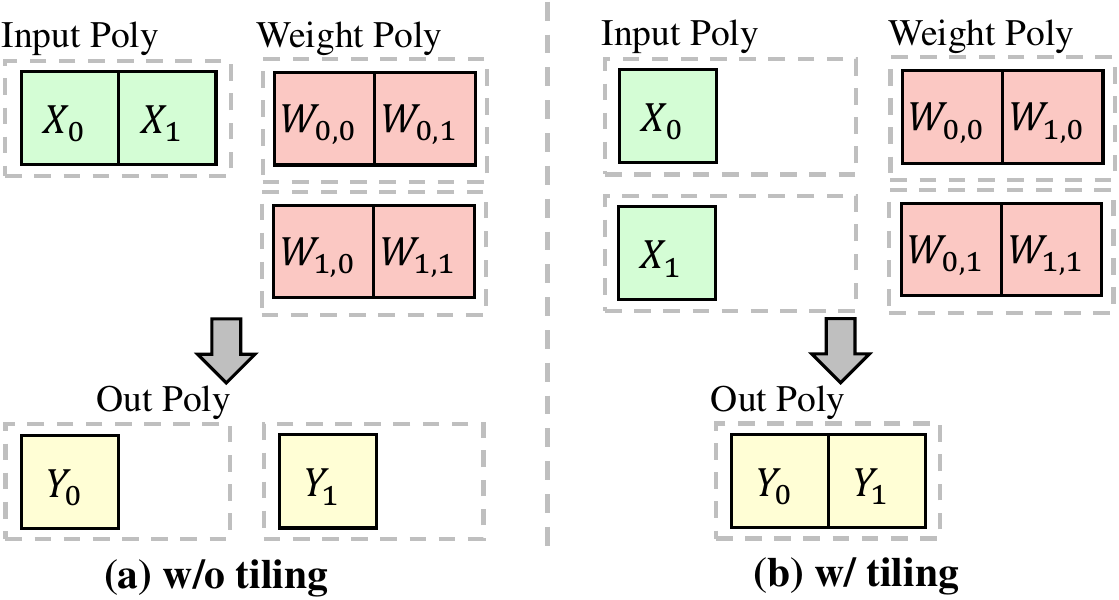}
    \caption{A toy example of operator tiling where each square means a channel and $W_{0,1}$ means the second channel of the first kernel of the filter.}\label{fig:toy_tile}
\end{figure}

Now we formulate the tiling optimization problem as a nonlinear programming problem. Note that in Equation~\ref{eq:comm_cost1}, only $C_x,C_w$ needs to be determined, with all
other parameters are known before execution. Moreover, we have two constraints: \textbf{\underline{1)}} polynomial coefficient encoding should not exceed the degree $N$, 
and \textbf{\underline{2)}} $C_x$ must be a positive integer multiple of $C_w$ to avoid doubling of the number of multiplications. Consequently, the tiling optimization problem can be formulated as follows:
\begin{align*}
    \mathrm{min}\ \ \ \ &\lceil \frac{C}{C_x}\rceil\times Nq+(\lceil \frac{K}{C_w}\rceil\times N+HWK)q\\
    \text{s.t.} \quad &C_xHW\le N \\ 
    &C_x=kC_w\text{ where }k \in \mathbb{Z^+}
\end{align*}
This is a solvable nonlinear programming problem with a relatively small solution space. With $C_x\le \frac{N}{HW}$ and $C_x=kC_w$, the solution space
is at most $\frac{N}{HW} \times  \frac{N}{HW}$, enabling direct solution through a search algorithm with a complexity of $O((\frac{N}{HW})^2)$. An example is illustrated in Figure~\ref{fig:toy_tile}, after tiling, we reduce the size of output polynomials.
\section{Experimental Results}
\subsection{Experimental Setup}
\method~is built on top of the SEAL library \cite{sealcrypto}, the EMP
toolkit~\cite{emp-toolkit} and OpenCheetah~\cite{huang2022cheetah} in C++. We also use the Ezpc
library~\cite{chandran2017ezpc} to evaluate
CrypTFlow2~\cite{rathee2020cryptflow2}. 
Consistent with
~\cite{huang2022cheetah,Shen_Dong_Fang_Shi_Wang_Pan_Shi_ABNN2_2022,mohassel2017secureml}, we simulate
a LAN and WAN network setting via Linux Traffic Control, where the bandwidth is 384 MBps and 9 MBps, respectively.
All the following experiments are performed on machines with 2.2 GHz Intel Xeon
CPU. Following~\cite{hao2022iron,huang2022cheetah}, we set
$N=4096$.
For network quantization, we use PyTorch framework for quantization-aware training, and for the inference accuracy evaluation.
We follow~\cite{bhalgat2020lsq+}'s setting to train ResNet18 on CIFAR-100 and Tinyimagenet with different bit width, including W8A8, W4A4, W3A3, W2A3 (WxAy means x-bit weights and y-bit activation).
We follow the setting of~\cite{yamamoto2021LCQ} to train ResNet50 on ImageNet with bit width W8A8, W4A4, W3A3 and W2A2. What's more, we use strategy3 for \method~ to determine the accumulation bit width. The more training details are shown in Appendix~\ref{app:exp_setting}. 


\subsection{Micro-Benchmark Evaluation}\label{subsec:micro}

\begin{figure}[!tb]
    \centering
    \includegraphics[width=1.0\linewidth]{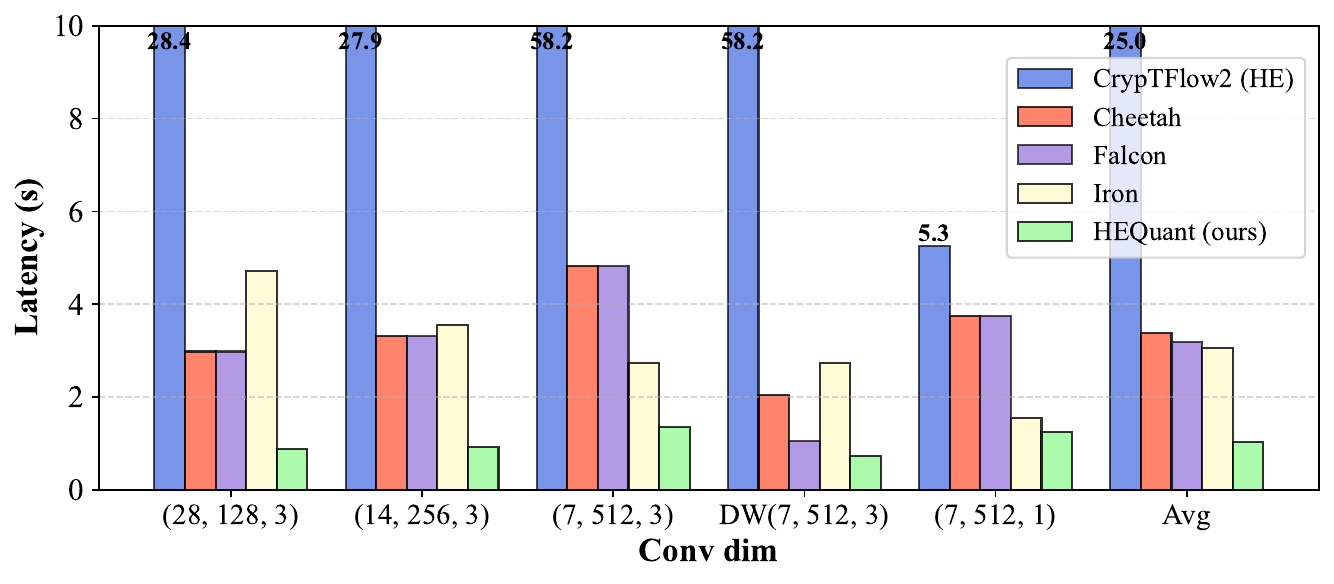}
    \caption{Latency comparison of
    convolutions of different dimensions (Conv dim represents the input feature 
    height, channels, and kernel size). DW means depthwise convolution and Avg means the average.}\label{fig:exp_conv}
\end{figure}

\begin{figure*}[!tb]
    \centering
    \includegraphics[width=0.9\linewidth]{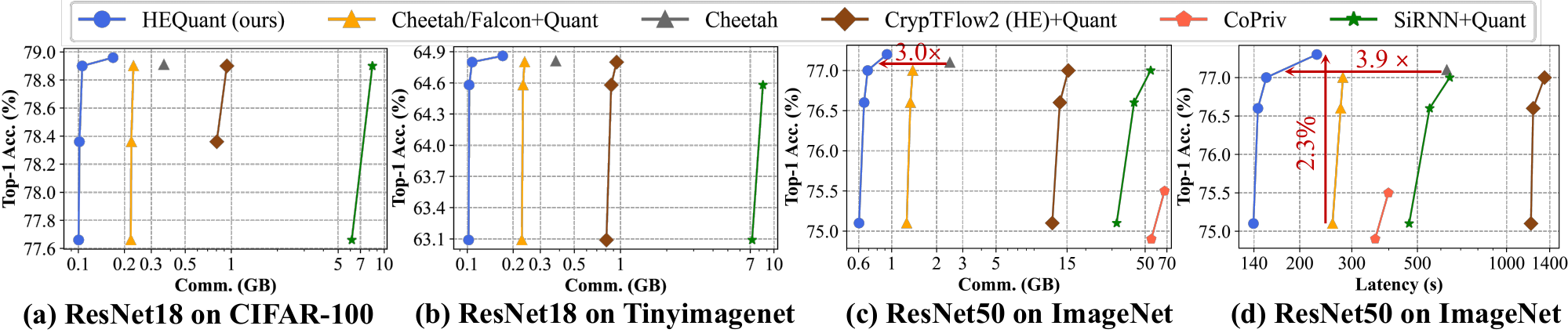}
    \caption{Comparison with prior-art 2PC frameworks on three benchmarks. }\label{fig:exp_protocol}
    \vspace{-10pt}
\end{figure*}

\begin{figure*}[!tb]
    \centering
    \includegraphics[width=0.9\linewidth]{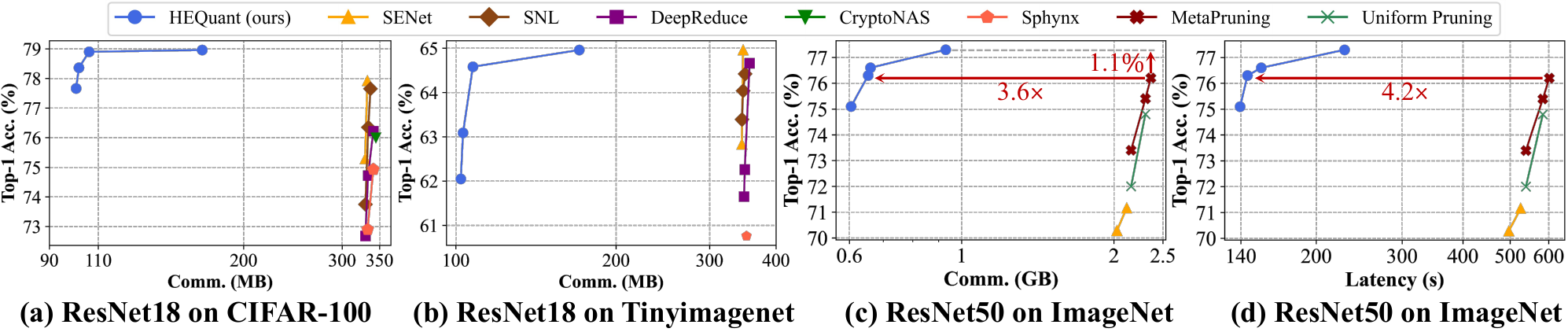}
    \caption{Comparison with prior-art network optimization algorithms on three benchmarks. }\label{fig:exp_network}
    \vspace{-10pt}
\end{figure*}

We compare the performance of \method~with CrypTFlow2, Iron, Cheetah, and Falcon for single convolution and depthwise convolution with different dimensions.
We determine the accumulation bit width based on W4A4 for~\method. 
As shown in Figure~\ref{fig:exp_conv},
\method~achieves on average $3.0\times$ latency reduction compared to SOTA Iron.
Note that due to the use of SIMD-based HE in CrypTFlow2, there is an extremely high latency\cite{huang2022cheetah}.
In comparison, \method~achieves a notable latency reduction of $31\times$. We also provide the results of communication in Appendix~\ref{app:micro}. 

\subsection{End-to-End Inference Evaluation}


\noindent \textbf{Benchmark with prior-art 2PC frameworks}
We first benchmark \method~with prior-art 2PC frameworks which focus on secure protocols' improvement, including Cheetah, Falcon, CoPriv, etc~\cite{huang2022cheetah,xu2023falcon,zeng2023copriv,rathee2021sirnn,rathee2020cryptflow2}. We conduct the end-to-end private inference in our quantized networks with different bit widths. We draw the
Pareto curve of accuracy and communication/latency (LAN) in Figure~\ref{fig:exp_protocol}. We put results of latency under WAN in Appendix~\ref{app:exp_la}. 

\textbf{Result and analysis} From Figure~\ref{fig:exp_protocol}, we make the following observations: \textbf{\underline{1)}}  \method~achieves SOTA pareto front in all three benchmarks. More specifically, compared with Cheetah, \method~achieves $3.0\times $ communication and $3.9\times$ latency reduction in ImageNet.
Compared with cheetah with the same quantization bit width, \method~achieves $ 2.0\sim 2.3\times$ communication reduction and $1.4\sim 1.9\times$ latency reduction in three benchmarks, respectively. Moreover, \method~achieves $2.3\%$ higher accuracy with less latency in ImageNet. \textbf{\underline{2)}} Compared with OT-based frameworks, SiRNN and CoPriv \cite{zeng2023copriv}, \method~demonstrates strong superiority and outperform them with $52\sim 83\times$ and $91\sim 111\times$ communication reduction, respectively. This translates to a latency improvement of an order of magnitude under WAN, as shown in Appendix~\ref{app:exp_la}. Since Falcon is specifically optimized for MobileNet with depthwise convolution, we compare \method~ with Falcon on MobileNetV2 in Appendix~\ref{app:falcon} and demonstrate $2.5\times$ communication reduction.


\noindent \textbf{Benchmark with prior-art network optimization algorithms}
Since \method~requires to quantize the network, 
we also benchmark \method~with other different network optimization algorithms, including ReLU-optimized networks (SENet, SNL, DeepReduce, etc) and SOTA network pruning methods (MetaPruning and uniform pruning)~\cite{kundu2023SENet,cho2022SNL,jha2021deepreduce,liu2019metapruning}. We apply those algorithms to the 2PC framework Cheetah and draw the Pareto curve in Figure~\ref{fig:exp_network}.

\textbf{Result and analysis} From the results in Figure~\ref{fig:exp_network}, we observe: \textbf{\underline{1)}} Compared with ReLU-optimized networks, \method~achieves over $3\times$ communication by both reducing the linear and non-linear layers' communication; \textbf{\underline{2)}} Compared with network pruning method MetaPruning, \method~achieves $3.6\times$ communication reduction and $4.2\times$ latency reduction as well as $1.1\%$ higher accuracy in ImageNet.

\subsection{Ablation Study}\label{subsec:ablation}
\noindent \textbf{Accumulation bit width reduction strategies}
In Table~\ref{table:exp_ablation_bit_extend}, we show the communication of different accumulation bit width reduction strategies.
We find that even with Strategy 1, \method~still achieves $3.2\times$ communication reduction compared with Cheetah.
With Strategy 3, \method~achieves on average $3.9\times$ communication reduction over Cheetah.

\begin{table}[!tb]
    \centering
    \caption{Communication (MB) comparison with different accumulation bit width reduction strategies.}
    \label{table:exp_ablation_bit_extend}
    \resizebox{\linewidth}{!}{
    \begin{tabular}{|c|c|c|c|c|}
    \hline
    \multirow{1}{*}{\textbf{Network and Dataset}}&\multicolumn{1}{c|}{\textbf{Cheetah}}
     & \multicolumn{1}{c|}{\textbf{\makecell{\method \\(Strategy1)}}}&
      \multicolumn{1}{c|}{\textbf{\makecell{\method \\(Strategy2)}}}&
      \multicolumn{1}{c|}{\textbf{\makecell{\method \\(Strategy3)}}}\\
    \hline
    ResNet18 (Cifar100)& 362&153 $(2.4\times \downarrow)$&141 $(2.6\times \downarrow)$&102 $(3.6\times \downarrow)$ \\
    \hline
    MobileNetV2 (Tinyimagenet)& 608&132 $(4.6\times \downarrow)$&123 $(4.9\times \downarrow)$&114 $(5.3\times \downarrow)$\\
    \hline
    ResNet50 (Imagenet)& 2454&771 $(3.2\times \downarrow)$& 689 $(3.7\times \downarrow)$& 654 $(3.8\times \downarrow)$\\
    \hline
    Average & 1141&352 $(3.24\times \downarrow)$&318 $(3.59\times \downarrow)$&290 $(3.94\times \downarrow)$\\
    \hline
    \end{tabular}
    }
\end{table}
\begin{figure}[!tb]
    \centering
    \includegraphics[width=0.78\linewidth]{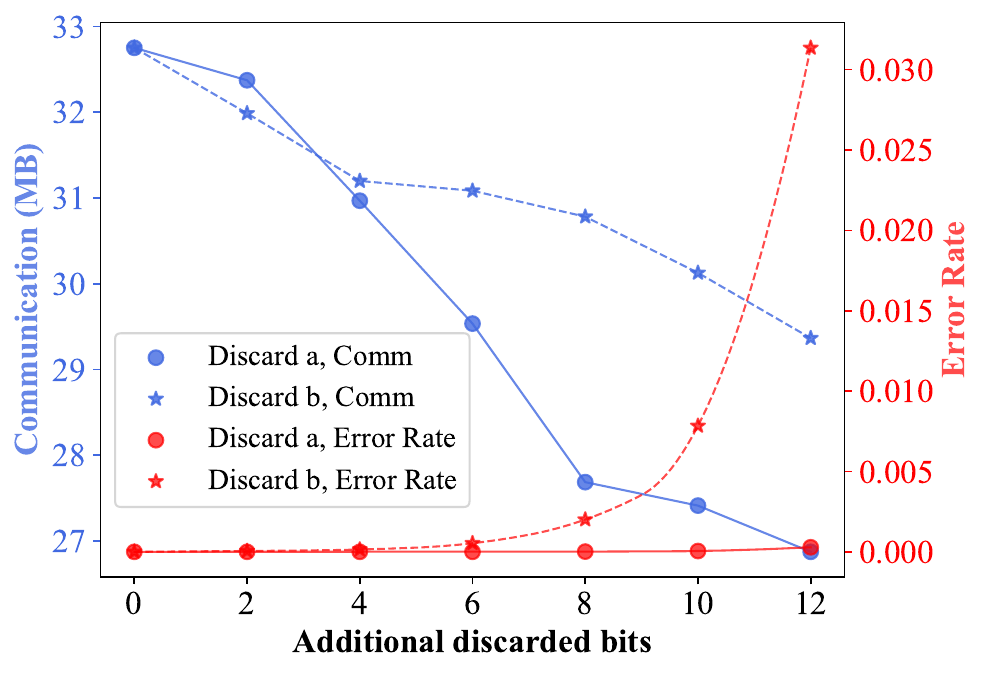}
    \caption{Communication and output error rate for different discarded bits where ``Discard a'' represents fixing the bit width of b and discarding more bits of $\bm{a}$.  }
    \label{fig:ablation_hack}
    \vspace{-10pt}
\end{figure}

\noindent \textbf{Output coefficient optimization analysis }
To show the trade-off between communication and error rate in output coefficient optimization,
we change the number of discarded bits and profile the communication as well
as the 1-bit error rate for a convolution from ResNet50.
We start with Cheetah
and adopt two strategies. The first ``Discard a'' means we fix the bit width of $b$ and discard more bits of $\bm{a}$ while ``Discard b'' means we discard more bits of $b$. As shown in Figure~\ref{fig:ablation_hack}, \textbf{\underline{1)}} the communication cost decreases with the number of discarded bits, while the error rate increases; \textbf{\underline{2)}} the communication cost of ``Discard a'' is lower than that of ``Discard b'' with the same additional discarded bits due to larger dimension of $\bm{a}$. In addition, we find the error of ``Discard a'' is lower, and hence, we choose ``Discard a'' as our strategy and limit the error rate below 0.0005.
\begin{figure}[!tb]
    \centering
    \includegraphics[width=0.9\linewidth]{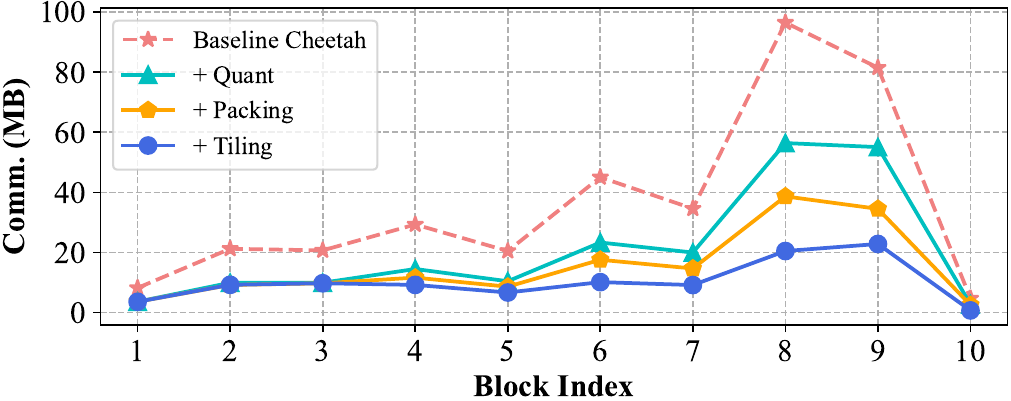}
    \caption{Block-wise visualization of communication on ResNet18 with different optimizations.}
    \label{fig:exp_ablation}
    \vspace{-10pt}
\end{figure}

\noindent \textbf{Block-wise comparison and ablation for different techniques }
To understand how different optimizations
help improve communication efficiency, we add our proposed optimizations step by step and visualize the block-wise communication reduction
on ResNet18 and present the results in Figure~\ref{fig:exp_ablation}. We can see that \textbf{\underline{1)}} both packing and tiling are effective and help further reduce the communication after quantization; \textbf{\underline{2)}} \method~achieves more communication reduction in deeper layers of the network, as
$H,W$ becomes smaller and $C $ becomes larger, resulting in a larger tiling
optimization space.
\section{Conclusion}
\label{sec:conclusion}
In this paper, we propose \method, the first 2PC framework that marries low-precision quantization and HE for efficient private inference. \method~features an intra-coefficient packing algorithm and a quantization-aware tiling algorithm to further reduce the communication for low-precision DNNs. Compared with prior-art 2PC frameworks, e.g. Cheetah, \method~achieves $3.0\times$ communication and $3.9\times$ latency reduction, respectively. Compared with the prior-art network optimization algorithms, \method~can achieve over $3.1\times$ communication reduction with higher accuracy.

\appendix
\clearpage
\setcounter{page}{1}
\maketitlesupplementary

\section{Network quantization}\label{app:quantization}
Quantization is one of the most impactful ways to decrease the computational time and energy consumption of
neural networks. In neural network quantization, the weights and activation tensors
are stored in lower bit precision than the 16 or 32-bit precision they are usually trained in. When moving from 32 to 8 bits, the memory overhead of storing tensors decreases by a factor of 4 while the
computational cost for matrix multiplication reduces quadratically by a factor of 16. Neural networks
have been shown to be robust to quantization~\cite{nagel2021white_quant,gholami2022survey_quant}, meaning they can be quantized to lower bit-widths
with a relatively small impact on the network's accuracy. Neural network quantization is an essential step in the model efficiency pipeline
for any practical use-case of deep learning~\cite{krishnamoorthi2018quantizing_white,bhalgat2020lsq+,yamamoto2021LCQ,liu2022nonuniform}.

\section{Output coefficient optimization}\label{app:hack}
Recall from Section~\ref{subsec:intra}, a $p$-bit output $m$ is encrypted into a scalar $b$ and a polynomial $\bm{a}$.
During decryption, we compute $m=\lfloor (b+\bm{a}^\text{T}\bm{s})P/Q\rfloor$ to reconstruct the output $m$ and we let $m_e=b+\bm{a}^\text{T}\bm{s}$. We observe that discarding the lower bits of $b$ and $\bm{a}$ before ciphertext transmission can reduce the communication cost and doesn't affect the correctness of decryption.

Formally, we write the high-end and low-end parts of $b$ and $\bm{a}$ as $ b = b_H2^{l_b} +b_L$ and $\bm{a} = \bm{a}_H2^{l_a} +\bm{a}_L$ such that $0\le b_L< 2^{l_b},0\le \bm{a}_L< 2^{l_a}$. Then $m_e=b_L+\bm{a}_L^\text{T}\bm{s}+b_H2^{l_b}+\bm{a}_H^\text{T}\bm{s}2^{l_a}$. And we let $e_L=b_L+\bm{a}_L^\text{T}\bm{s}$. If we discard the low-end parts of $b$ and $\bm{a}$, after decryption, we will get $\lfloor (m_e-e_L)P/Q\rceil=m-\lfloor e_L P/Q\rceil+\lfloor eP/Q\rceil$, where $\lfloor eP/Q\rceil$ will round to 0. Remind that only the upper half of the bits of $m$ are useful, so our goal is to ensure $m\gg \frac{p}{2}\equiv (m-\lfloor e_L P/Q\rceil)\gg \frac{p}{2}$ where $\gg$ means right shift. We first approximate it by $(m-\lfloor e_L P/Q\rceil)\gg \frac{p}{2}\approx (m\gg \frac{p}{2})-(\lfloor e_L P/Q\rceil \gg \frac{p}{2})$. Therefore, we need to ensure $\lfloor e_L P/Q\rceil \gg \frac{p}{2}=\lfloor (b_L+\bm{a}_L^\text{T}\bm{s}) P/Q\rceil \gg \frac{p}{2}=0$. The key here is to find out an upper bound for $\bm{a}_L^\text{T}\bm{s}$. Since the secret vector $\bm{s}$ distribute uniformly in $\{0, ±1\}^N$ and $\bm{a}_L$ distributes uniformly in $[0,2^{l_a})$. Then the variance of $\bm{a}_L^\text{T}\bm{s}$ is $\mathrm{Var}\approx \frac{2N}{9}(2^{l_a})^2$. Same with Cheetah~\cite{huang2022cheetah}, we use $7\sqrt{\mathrm{Var}}$ as a high-probability upper bound $(\approx 1-2^{-38.5})$ on the value $\bm{a}_L^\text{T}\bm{s}$. Finally, the constraint is $(7\sqrt{\frac{2N}{9}}2^{l_a}+2^{l_b})P/Q<2^{\frac{p}{2}}$ and we set $l_b=l^\prime+\frac{p}{2}-1,\ l_a=l^\prime+\frac{p}{2}-1-\lfloor \log_2(7\sqrt{\frac{2N}{9}})\rfloor$ to meet the constraint, where $l^\prime = \lfloor \log_2(Q/P)\rceil$. 

However, the approximation of $(m-\lfloor e_L P/Q\rceil)\gg \frac{p}{2}\approx (m\gg \frac{p}{2})-(\lfloor e_L P/Q\rceil \gg \frac{p}{2})$ may cause least significant bit (LSB) error. To be more specific, there will be a $1/(2^{q-p/2-\log_2(e_L)})$ probability of causing a 1-bit LSB error. Therefore, we decrease $e_L$ with several tricks to reduce the probability of LSB error. The first trick is to decrease the number of discarded bits of $b$ rather than $\bm{a}$ because $\bm{a}$ accounts for larger communication. Moreover, after each linear layer, we need to truncate the output to low bit width $b_a$ where the output will be divided by a scaling factor. That means more low-end bits of the output are useless and can help us reduce the 1-bit error rate. Ultimately, we make the error rate lower than 0.0005 by skipping the low $l^\prime-1$ of $b$ and the low $l^\prime +\frac{p}{2}-1-\sigma$ of $\bm{a}$ where $\sigma=\lfloor \log_2(7\sqrt{\frac{2N}{9}})\rfloor$.

\section{Visualization of different accumulation bit width reduction strategies}\label{app:accum_reduce}
\begin{figure}[!tb]
    \centering
    \includegraphics[width=1.0\linewidth]{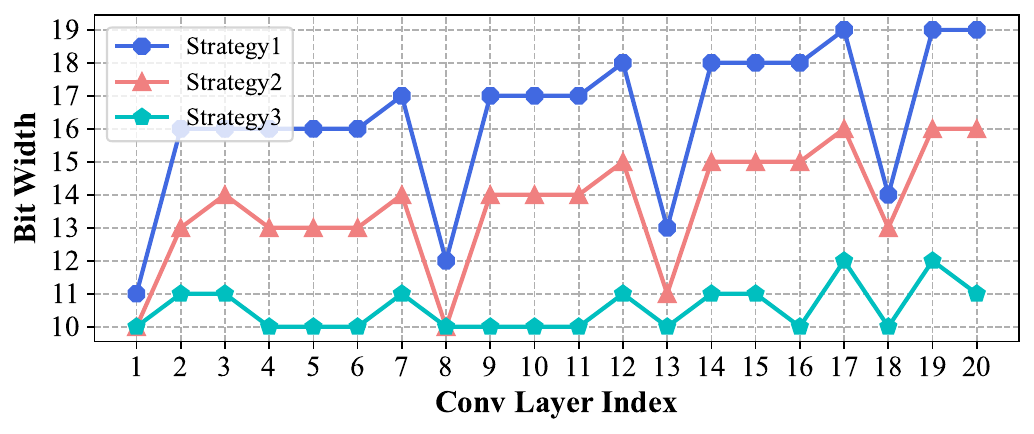}
    \caption{Layer-wise accumulation bit width in ResNet32 with 4-bit for both weight and activation.}\label{fig:ablation_bit_extend}
\end{figure}
Figure~\ref{fig:ablation_bit_extend} visualizes the layer-wise accumulation bit width in a ResNet18 network trained on the CIFAR-100 dataset with different strategies proposed in Section~\ref{subsec:intra}.

\section{Experiment setting}\label{app:exp_setting}
\subsection{Network architecture}\label{app:exp_arch}
\begin{table*}[!tb]
    \caption{\method~evaluation benchmarks.}
    \label{tab:benchmarks}
    \centering
    \resizebox{0.9\linewidth}{!}{
    \begin{tabular}{cccccc}
        \toprule
        Model  &  Layers & \# Params (M) & MACs (G) & Dataset \\
        \hline
        ResNet18  & 20 Conv, 1 FC, 1 AP, 17 ReLU & 11.22 & 0.558 & CIFAR-100/Tinyimagenet \\
        ResNet50 &  49 CONV, 1 FC, 1, MP, 1 AP, 49 ReLU & 2.5 & 4.1 & Imagenet \\
        \bottomrule
    \end{tabular}
    }
\end{table*}
We evaluate and compare \method~with prior-art methods on
three networks on three datasets, and the details of which are shown in
Table~\ref{tab:benchmarks}.

\subsection{Training details}\label{app:exp_train}
We train ResNet18 following the training scheme in~\cite{bhalgat2020lsq+} to use real-valued PyTorch pre-trained models as initialization for corresponding quantized networks. We use Adam optimizer \cite{Kingma_Ba_2014_adam} with a cosine learning rate decay scheduler. The initial learning rate is set to $55e-5$ for weight parameters and batch size is set to $256$. We set weight decay to be $6e-2$. The models are trained for $400$ epochs \textbf{without} a knowledge distillation scheme. We adopt the basic data augmentation as ResNet~\cite{He_Zhang_Ren_Sun_2016resnet}. 
Training images are randomly resized and cropped to $32\times 32$ pixels ($64\times 64$ pixels for Tinyimagenet) and randomly flipped horizontally. Test images are center-cropped to the same resolution.

For ResNet50, we follow~\cite{yamamoto2021LCQ} with the same training settings. Specifically, with an initial learning rate of 0.1 for the weights and an initial learning rate of 0.01 for the clipping and companding parameters, the models were trained over 120 epochs with a mini-batch size of 512 for ResNet50. In addition, we applied a warm-up method for the first 5 epochs and increased the learning rate linearly from $10^{-4}$ to the initial value. The weight decay was set to $4 \times 10^{-5}$. The training images were resized, cropped to $224 \times 224$ pixels and randomly flipped horizontally. The test images were center-cropped to $224 \times 224$ pixels.

\section{Experimental results of latency}\label{app:exp_la}

\begin{figure*}[tb]
    \centering
    \includegraphics[width=1.0\linewidth]{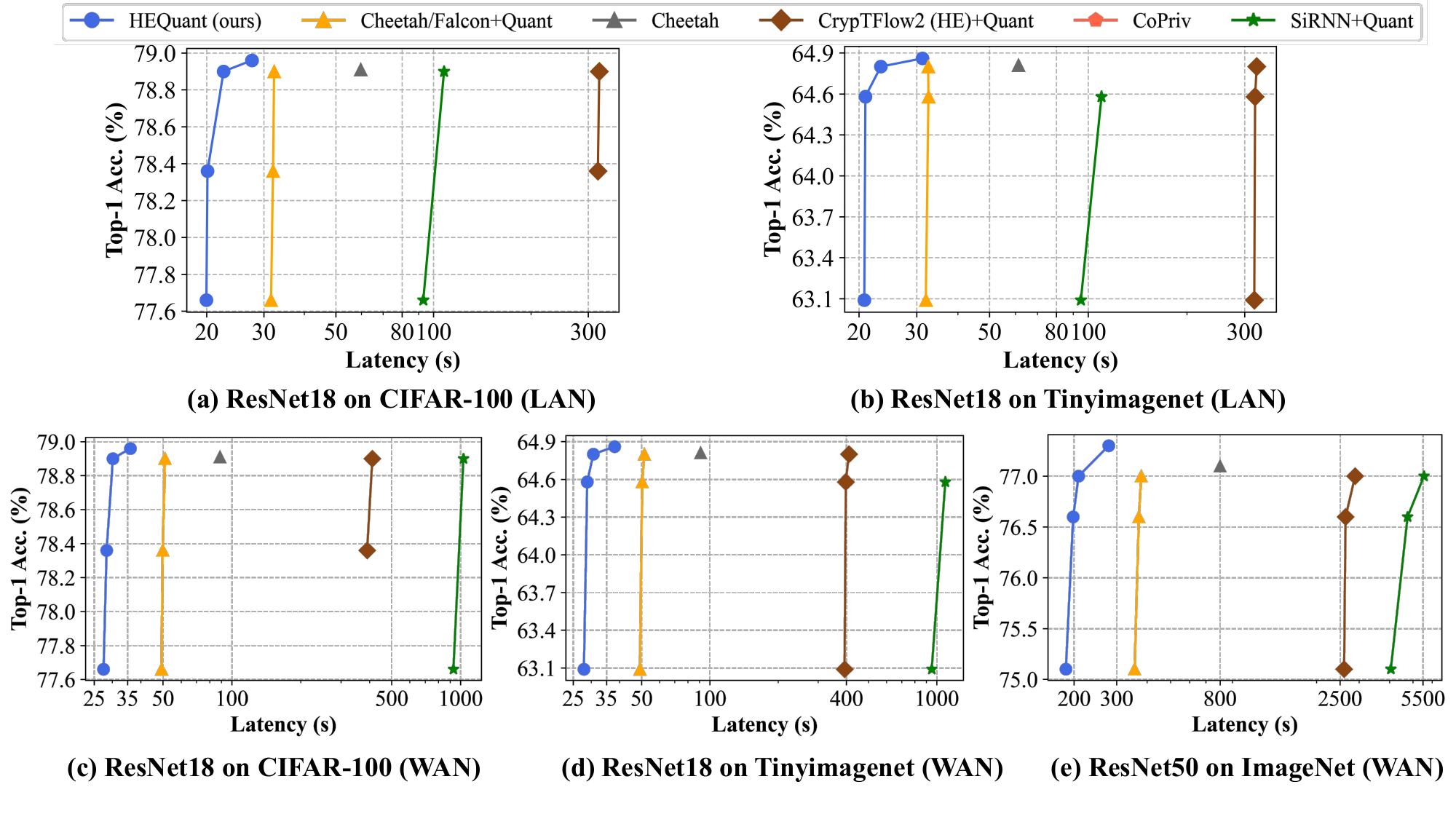}
    \caption{Latency comparison with prior-art 2PC frameworks on three benchmarks under LAN and WAN.}\label{fig:la_pro}
\end{figure*}

\begin{figure*}[tb]
    \centering
    \includegraphics[width=1.0\linewidth]{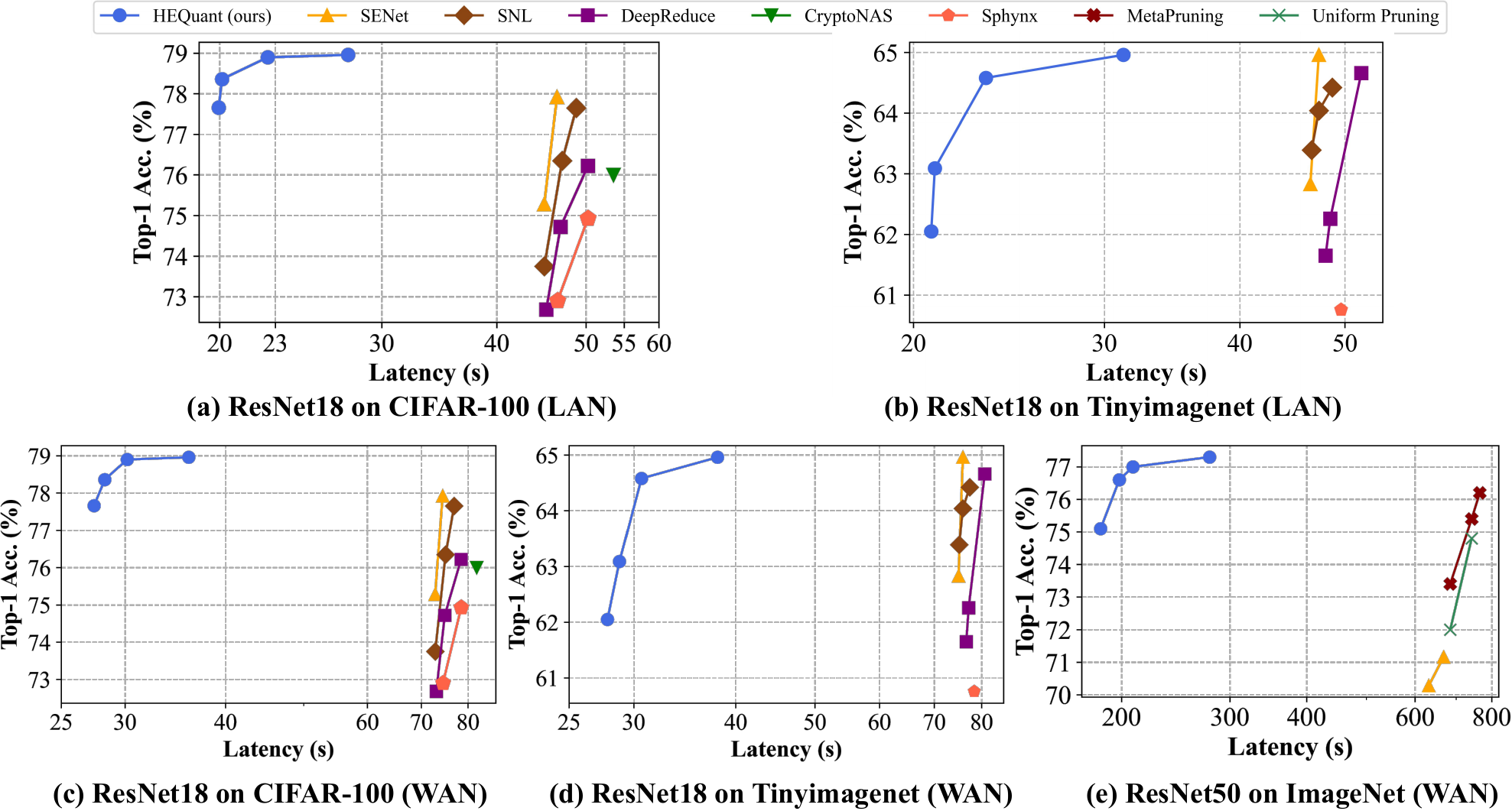}
    \caption{Latency comparison with prior-art network optimization algorithms on three benchmarks under LAN and WAN.}\label{fig:la_network}
    \vspace{14pt}
\end{figure*}

Figure~\ref{fig:la_pro} and Figure~\ref{fig:la_network} show the latency comparison with prior-art 2PC frameworks and network optimization algorithms on three benchmarks under LAN and WAN. We can see that: \textbf{\underline{1)}}: \method~achieves over $3\times$ latency reduction compared with prior-art 2PC frameworks and over $2.5\times$ latency reduction compared with prior-art network optimization algorithms; \textbf{\underline{2)}} Under WAN, \method~demonstrates more significant latency reduction than under LAN due to the large communication reduction achieves by \method.

\section{Communication comparison on Micro-Benchmark}\label{app:micro}

\begin{figure}[tb]
    \centering
    \includegraphics[width=1.0\linewidth]{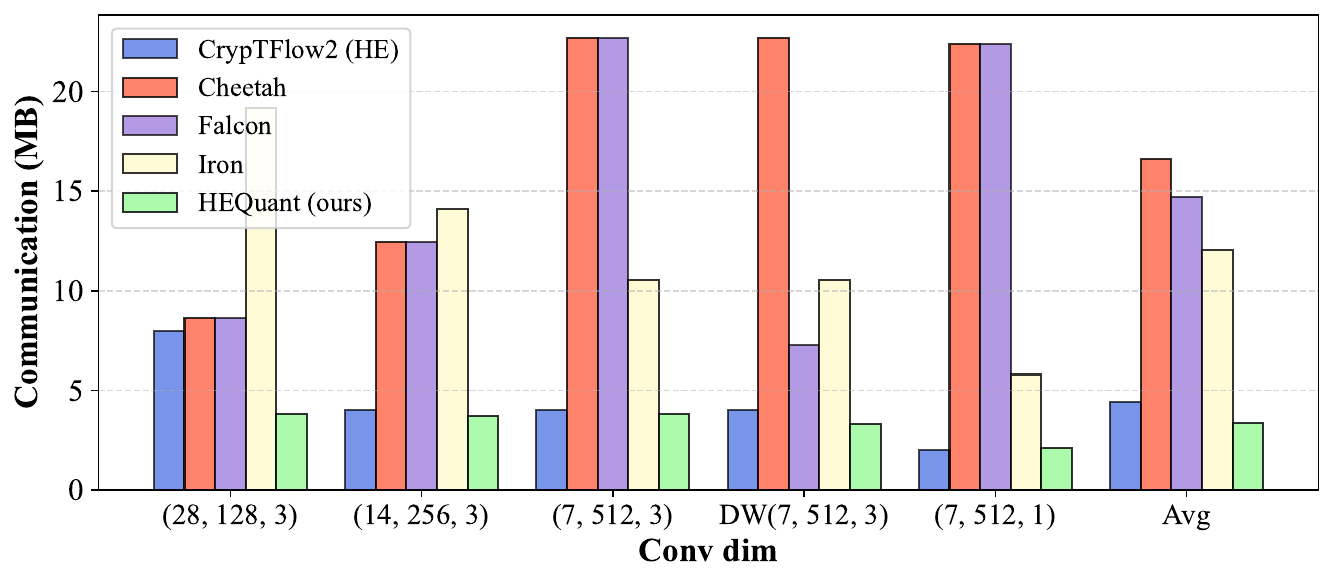}
    \caption{Communication comparison of
    convolutions of different dimensions (Conv dim represents the input feature 
    height, channels, and kernel size). DW means depthwise convolution and Avg means the average.}\label{fig:micro_conv_comm}
\end{figure}

Same with Section~\ref{subsec:micro}, we compare the communication cost of \method~with CrypTFlow2, Iron, Cheetah, and Falcon for single convolution and depthwise convolution with different dimensions.
As shown in Figure~\ref{fig:micro_conv_comm},
\method~achieves on average $3.0\times$ communication reduction compared to SOTA Iron.
Note that the communication cost of CrypTFlow2 is relatively low. However, CrypTFlow2 uses SIME-based HE and suffers from extremely high latency as shown in Section~\ref{subsec:micro}.

\section{Comparison with Falcon}\label{app:falcon}
We compare \method~with Falcon on MobileNetV2 in Table~\ref{table:exp_falcon}. Though Falcon is optimized specifically for MobileNetV2, \method~still achieves $2.5\times$ communication and $1.9\times$ latency reduction.
\begin{table}[!tb]
    \centering
    \caption{Comparison with Falon on MobileNetV2 on the TinyImagenet dataset with W4A4.}\label{table:exp_falcon}
    \resizebox{0.70\linewidth}{!}{
    \begin{tabular}{c|c|c}
    \toprule
    \textbf{MobileNetV2}&\textbf{Comm. (MB)}& \textbf{Latency (s)}\\
    \hline
    Cheetah& 561.7 $(4.8\times)$&86.3 $(3.0\times)$\\
    \hline
    Falcon& 296.9 $(2.5\times)$&55.9 $(1.9\times)$\\
    \hline
    \rowcolor{Gray}
    \method& 118.3 &29.0\\
    \bottomrule
    \end{tabular}
    }
\end{table}


\newpage
\newpage
\newpage
{
    \small
    \bibliographystyle{ieeenat_fullname}
    \bibliography{main}
}
\end{document}